# Algorithms for Closed Under
# Rational Behavior (CURB) Sets

**Michael Benisch**　　　　　　　　　　　　　　　　MBENISCH@CS.CMU.EDU
**George B. Davis**　　　　　　　　　　　　　　　　　　　GBD@CS.CMU.EDU
**Tuomas Sandholm**　　　　　　　　　　　　　　SANDHOLM@CS.CMU.EDU
*School of Computer Science*
*Carnegie Mellon University*
*5000 Forbes Ave*
*Pittsburgh, PA 15213 USA*

## Abstract

We provide a series of algorithms demonstrating that solutions according to the fundamental game-theoretic solution concept of closed under rational behavior (CURB) sets in two-player, normal-form games can be computed in polynomial time (we also discuss extensions to $n$-player games). First, we describe an algorithm that identifies all of a player's best responses conditioned on the belief that the other player will play from within a given subset of its strategy space. This algorithm serves as a subroutine in a series of polynomial-time algorithms for finding all minimal CURB sets, one minimal CURB set, and the smallest minimal CURB set in a game. We then show that the complexity of finding a Nash equilibrium can be exponential only in the size of a game's smallest CURB set. Related to this, we show that the smallest CURB set can be an arbitrarily small portion of the game, but it can also be arbitrarily larger than the supports of its only enclosed Nash equilibrium. We test our algorithms empirically and find that most commonly studied academic games tend to have either very large or very small minimal CURB sets.

## 1. Introduction

For noncooperative multi-agent settings, game-theoretic solution concepts help players choose strategies, help modelers predict outcomes, and help mechanism designers guarantee properties of the systems they create. Significant attention has been given to algorithms for computing solutions according to concepts of subgame perfect Nash equilibrium (e.g., minimax search and $\alpha$-$\beta$-pruning), Nash equilibrium (Lemke & Howson, 1964; Porter, Nudelman, & Shoham, 2004; Sandholm, Gilpin, & Conitzer, 2005), correlated equilibrium (Gilboa & Zemel, 1989), iterative dominance (Knuth, Papadimitriou, & Tsitsiklis, 1988; Conitzer & Sandholm, 2005a), and other related concepts (Conitzer & Sandholm, 2005b).

The Nash equilibrium concept, under which each player weakly prefers its strategy as long as the other players do not deviate from theirs, remains the most important point-valued game-theoretic solution concept. However, it has been shown that, even in two-player games with binary utilities, computing a single Nash equilibrium is PPAD-complete (Chen & Deng, 2006; Abbott, Kane, & Valiant, 2005), suggesting that no algorithms exist for computing these equilibria in worst-case polynomial time (Daskalakis, Goldberg, & Papadimitriou, 2009).

There are other fundamental solution concepts that have some known advantages over Nash equilibria, and—as we will show—solutions according to some of these concepts can be found in polynomial time, even in the worst case. Specifically, we will study the fundamental concept of *closed under rational behavior* (CURB) strategy sets in two-player, normal-form games. A game can have multiple Nash equilibria, but each of those is a point, or a single (potentially mixed) strategy for each player. In contrast, a CURB set can contain multiple strategies for each player, and it is stable so long as players choose any (potentially mixed) strategies from within the set.

CURB sets are based on the notion of *rationalizability*, which was introduced by Pearce (1984) and Bernheim (1984). Rationalizability is, by now, a widely known, robust game-theoretic solution concept and has been used to study various applications, such as first-price auctions (Battigalli & Siniscalchi, 2003). Its main insight is that rationality restricts players from ever playing strategies that are not best responses given the beliefs they hold about their opponents. Strategies that are not best responses to a set of consistent beliefs about opposing strategies are said not to be *rationalizable*. In two-player games, the process of iteratively eliminating strategies that are *dominated*, that is strategies that are not best responses to any opponent strategy, captures the concept of rationalizability. It emulates the players' assumptions that an opponent will never play a strategy that is not a best response to one of the player's own remaining strategies (Pearce, 1984).

The set of all players' rationalizable strategies has the property that no player's best response to any pure or mixed strategy inside the set lies outside the set – in other words, the set is CURB. However, this CURB set may also have CURB subsets, which demonstrates how CURB sets extend the notion of rationalizability. Basu and Weibull (1991) introduced the notion of a *minimal* CURB set, or a CURB set that does not contain any CURB subsets, and proved that each minimal CURB set is guaranteed to contain the supports of at least one Nash equilibrium.

The minimal CURB set solution concept has since been motivated from several perspectives in academic literature, including the following:

- Mixed-strategy Nash equilibria (the type guaranteed to exist in every game) can be highly unstable, because a player may be indifferent between some of its strategies. Strict Nash equilibria, where players strictly prefer their strategies in equilibrium, are a more stable alternative, but they are not guaranteed to exist. Minimal CURB sets always exist and have been referred to as the "nearest set-valued generalization of strict Nash equilibria," since they are the smallest sets of strategies that include all ways of choosing among the indifferences of an equilibrium (Basu & Weibull, 1991).

- A CURB set can be viewed as a subspace of strategies within which any best-response dynamic (even a best-response dynamic of mixed strategies) will stay. Thus, CURB sets have been used as a solution concept to describe the strategy subspace where iteratively adapting agents will eventually settle (Hurkens, 1995).

- Voorneveld *et.al.* enumerated a number of other properties of minimal CURB sets that illustrate the stability of set-based solution concepts over point-valued concepts, such as Nash equilibria (Voorneveld, Kets, & Norde, 2005).

In order for a solution concept to be operational, it must also be accompanied by algorithms for applying it. Finding minimal CURB sets has been previously considered challenging (Pruzhansky, 2003), and, prior to this work on CURB sets, little had been done on the problem from a computational standpoint. To our knowledge, the only such work to predate ours was that of Pruzhansky, which studied sequential games of perfect information. Such games are relatively simple, in that they contain exactly one minimal CURB set, and a straightforward algorithm can quickly find it by exploiting the sequential representation (Pruzhansky, 2003). In this paper, we present the first thorough computational treatment of CURB sets in general two-player games. We show that, in these settings, minimal CURB sets are actually easy to find: the time complexity is polynomial in the size of the game, even in the worst case.

The primary source of complexity for our algorithms is a linear programming-based subroutine for finding all of a player's best responses (i.e., utility-maximizing pure strategies) conditioned on the belief that the other player will play from within a given subset of its strategy space. This problem can be solved fast for two players, in this case it involves solving a simple linear feasibility program, but the mathematical program we use is of degree $p - 1$, where $p$ is the number of players, and with $p = 3$ the constraints are already quadratic. On the plus side, our CURB set algorithms only make a polynomial number of calls to this subroutine. If future research is able to identify polynomial-time algorithms for finding all of a player's best responses in $n$-player games, then our CURB set algorithms will also be polynomial time in those settings. Additionally, our algorithms have been useful as templates for the development of other algorithms to compute related solution concepts in $n$-player games (Brandt, Brill, Fischer, & Harrenstein, 2009; Jordan & Wellman, 2010).

The rest of the paper is organized as follows. We begin with some preliminaries on our notations and definitions. Next, we present and analyze our algorithm for finding conditional best responses, which serves as the main subroutine of our CURB set finding algorithms. We then present and analyze a family of polynomial-time algorithms for two-player, normal-form games that compute all of a game's minimal CURB sets, a single one of its minimal CURB sets, and its smallest minimal CURB set. Finally, we discuss additional applications of our results, including the potential for our CURB set algorithms to bound the theoretical complexity of finding Nash equilibria.

## 2. Preliminaries

We describe and analyze our algorithms in the classic game-theoretic setting of a two-player, normal-form game represented as a matrix with rows corresponding to the *pure strategies* (or actions) of one player, player $r$, and columns corresponding to the pure strategies of the other, player $c$. (For shorthand, we will often omit the term "pure" and refer to a pure strategy simply as a *strategy*.) As is typical in game theory, the players are assumed to be fully-rational, utility-maximizing agents.

Each row in the game matrix corresponds to a strategy, $s_r$, from the set of all player $r$'s strategies, $S_r$. Likewise, each column corresponds to a strategy, $s_c$, from the set of all of player $c$'s strategies, $S_c$. The cell corresponding to strategies $s_r$ and $s_c$ contains two entries, one indicating the real-valued utility of the row player when $s_r$ and $s_c$ are played, $u_r(s_r, s_c)$, and the other indicating that of the column player when those two strategies are played, $u_c(s_r, s_c)$. Using these entities, we also refer to a game, $\mathcal{G}$, as a tuple, $\mathcal{G} = \langle S_r, S_c, u_r, u_c \rangle$. The *size* of the game, which we refer to as $n$, is the total number of strategies it contains, $n = |S_r| + |S_c|$.

A *mixed strategy*, or *mixture*, is a probability distribution over pure strategies, or a function, $m_i$, that maps from each of player $i$'s pure strategies to a probability, $m_i : S_i \rightarrow [0, 1]$ and $\sum_{s \in S_i} m_i(s) = 1$. The *supports* of a mixture are all of the pure strategies in the mixture with non-zero probability. The set of all possible mixtures with supports in some set of strategies, $S_i$, is denoted $M(S_i)$, and can be thought of as a simplex with degree $|S_i| - 1$. A pure strategy can be represented as a *point-mass* mixture, that is a mixture with all of its probability mass on one strategy.

A *strategy profile* is a set of pure or mixed strategies, with one for each player. When a mixed-strategy profile is played, a player's utility is assumed to be its *expected utility*, which is given by summing that player's utility for each possible pure-strategy profile weighted by the profile's joint probability according to the mixtures, e.g., $u_r(m_r, m_c) = \sum_{s_r \in S_r} m_r(s_r) \sum_{s_c \in S_c} m_c(s_c) u_r(s_r, s_c)$. We occasionally use the notation $-i$ to refer to the player or players other than some player $i$. When used as a subscript on a strategy-related entity with more than two players, we intend for the $-i$ to refer to one instance of the entity per player (e.g., $m_{-i}$ is a mixed-strategy profile containing one mixture per player other than $i$).

Player $i$'s *best responses* to a mixed strategy of the other player(s), $m_{-i}$, are given by the function $\beta_i(m_{-i})$. These are the pure strategies that maximize player $i$'s utility if the other player(s) play $m_{-i}$.

For a set of the other player's pure strategies, $S_{-i}$, we define $\beta_i(S_{-i})$ to be a function that returns player $i$'s best responses to every mixture with supports in $S_{-i}$, $\beta_i(S_{-i}) = \bigcup_{m \in M(S_{-i})} \beta_i(m)$. In Section 3, we describe an algorithm for computing the pure strategies in $\beta_i(S_{-i})$ that serves as a subroutine in our CURB set algorithms, and we refer to the strategies it computes as *conditionally rational*. We define $\beta(S)$ (without a subscript $i$) as the union of the sets $\beta_i(S_{-i})$ for all players.

A *CURB set*, $S$, is formally defined as a set of pure strategies (with at least one strategy for each player) that contains the best responses to any mixture over itself: if $S$ is CURB and players believe that no strategy outside of $S$ will be played with positive probability by their opponents, then such strategies will indeed not be played by rational players. Using the notation above, a set, $S$, is CURB if $\beta(S) \subseteq S$. (The entire game is trivially CURB by this definition.) We refer to the number of strategies in a CURB set as its *size*. The *intersection* of two CURB sets, $S_1$ and $S_2$, is the set of strategies attained by taking the intersection of their strategy sets, $S_I = S_1 \cap S_2$. Two CURB sets *overlap* if they share a strategy (i.e., their intersection is non-empty).

A *Nash equilibrium* is a pure- or mixed-strategy profile, $\{m_r, m_c\}$, such that each player's strategy is at least as good as its best response to the other's, $u_r(m_r, m_c) = u_r(s_r^*, m_c)$ and $u_c(m_r, m_c) = u_c(m_r, s_c^*)$, where $s_r^* \in \beta_r(m_c)$ and $s_c^* \in \beta_c(m_r)$. A *strict Nash equilibrium* is a pure-strategy profile, $\{s_r, s_c\}$, where each player's strategy is its only best response to the other's, $\beta_r(s_c) = \{s_r\}$ and $\beta_c(s_r) = \{s_c\}$. A CURB set that contains only one strategy per player is also a pure-strategy Nash equilibrium.

## 3. Finding Conditional Best Responses

Finding all of a player's best responses conditioned on the belief that the other player will play from within some subset of its strategy space, is a problem of interest in its own right, and it plays a central role in our computation of CURB sets. This section describes a polynomial-time algorithm, `all_conditionally_rational`, for doing just that. The algorithm below is for the row player; the column player's variant is symmetric. The inputs to the algorithm are a set of row-player strategies to consider, $S_r$, a set of column-player strategies they may be played against, $S_c$, and the row player's utility function, $u_r$.

**function** `all_conditionally_rational`$(S_r, S_c, u_r)$

  $S_r^* \leftarrow \emptyset$

  **for each** row strategy, $s_r \in S_r$ **do**

   **if** there exists a feasible solution to the following linear feasibility program:

   **find** $p_{s_c}$ **such that**

$$\sum_{s_c \in S_c} p_{s_c} = 1 \tag{1}$$

$$\sum_{s_c \in S_c} p_{s_c} u_r(s_r, s_c) \geq \sum_{s_c \in S_c} p_{s_c} u_r(s_r', s_c) \quad \forall s_r' \in S_r \setminus s_r \tag{2}$$

$$p_{s_c} \geq 0 \quad \forall s_c \in S_c \tag{3}$$

   **then** $S_r^* \leftarrow S_r^* \cup s_r$

  **return** $S_r^*$

For each row strategy, $s_r \in S_r$, a linear feasibility program (LFP) (i.e., a linear program with no objective) is constructed to find a mixture of probabilities over column-player strategies, $p_{s_c}$, such that $s_r$ is the row player's best response. The constraints of the LFP ensure that the mixture is valid (sums to one), and that the row player's utility by playing $s_r$ against $p_{s_c}$ is greater than or equal to that of any other strategy. If the LFP has a feasible solution, $s_r$ is added to the set of best responses to be returned.

The computational complexity of the algorithms described in this paper depend on the total number of strategies in the game, $n$, and the complexity of solving an LFP with a number of variables and constraints bounded by $n$, which we will denote as LFP$(n)$. LFPs can be solved in low-order polynomial time, even in the worst case, and the fastest known algorithms for LFPs have better worst-case guarantees than the fastest known linear programming algorithms (Ye, 2006). In our experiments, we solve the LFP using the simplex algorithm, which has exponential worst-case time complexity, but is known to outperform polynomial-time linear programming algorithms in practice.

**Proposition 1.** *The* `all_conditionally_rational` *algorithm returns a player's best responses to every mixture over the input strategy set, and nothing else. Its worst-case time complexity is* $\Theta(n) \times LFP(n)$.

*Proof.* Since `all_conditionally_rational` runs this program on all strategies and includes them in the return set only if the LFP is feasible, it must be correct. Since the LFP is executed once for each strategy, and its size is bounded by $n$, `all_conditionally_rational` has complexity as shown. □

## 4. Finding CURB Sets

We now turn our attention to the problem of finding CURB sets. The algorithm below finds the smallest CURB set that contains a given seed strategy within a given subgame. (The returned set is not necessarily minimally CURB.) The algorithm repeatedly alternates between players, each time calling `all_conditionally_rational` to add the strategies necessary for maintaining the CURB property. If an iteration passes without any strategies being added, the algorithm has converged.

**function** `min_containing_CURB`$(s_r, \mathcal{G} = \langle S_r, S_c, u_r, u_c \rangle)$
    $S_r^* \leftarrow \{s_r\}$, $S_c^* \leftarrow \emptyset$
    **converged** $\leftarrow$ **false**
    **while** $\neg$**converged do**
      **converged** $\leftarrow$ **true**
      **for** $i \in \{r, c\}$ **do**
        $S_i' \leftarrow$ `all_conditionally_rational`$(S_i, S_{-i}^*, u_i)$
        **if** $S_i' \setminus S_i^* \neq \emptyset$ **then**
          **converged** $\leftarrow$ **false**
        $S_i^* \leftarrow S_i^* \cup S_i'$
    **return** $S_r^* \cup S_c^*$

It is worth noting that on the second-to-last line of the algorithm ($S_i^* \leftarrow S_i^* \cup S_i'$) it is necessary to merge the old strategies, $S_i^*$, with the newly identified strategies, $S_i'$, because $S_i'$ is not always a superset of $S_i^*$. If, instead, $S_i^*$ were replaced by $S_i'$, then it would be possible for the seed strategy to be eliminated during the algorithm's first iteration. For example, consider the following game.

|          | $s_{c_1}$ | $s_{c_2}$ |
|----------|-----------|-----------|
| $s_{r_1}$ | 1,1       | 0,0       |
| $s_{r_2}$ | 0,1       | 1,0       |

If strategy $s_{r_2}$ is used as a seed, then on the first iteration $S_r^*$ is initialized to $\{s_{r_2}\}$, $S_c^*$ is then set to $\{s_{c_1}\}$, and finally $S_r^*$ is set to $\{s_{r_1}\}$. Thus, without the merge the algorithm would output a subgame that does not contain the seed strategy.

**Proposition 2.** *The* `min_containing_CURB` *algorithm has worst-case runtime* $\Theta(n^2) \times LFP(n)$.

*Proof.* Every two calls made to `all_conditionally_rational` must add a strategy to the return set, or `min_containing_CURB` will terminate. Since at most $n$ strategies can be added this way, the complexity of `min_containing_CURB` is $\Theta(n^2) \times \text{LFP}(n)$. □

**Theorem 1.** *The* `min_containing_CURB` *algorithm is correct, that is, the returned set, $S^*$, is the smallest set of strategies that both 1) contains the given seed strategy, $s_r$, and 2) is CURB.*

*Proof.* The convergence of the algorithm implies that no strategies outside of $S^*$ are best responses to some mixture with supports in $S^*$. Therefore, $\beta(S^*) \subseteq S^*$, and $S^*$ is CURB.

To prove that $S^*$ is the smallest CURB set containing $s_r$, we can use induction on the strategies that are added.

**Base Case:** Initially, $S^*$ contains only $s_r$ and $\beta_c(s_r)$. At this point, $S^*$ is a subset of the the smallest CURB set containing $s_r$.

**Inductive Step:** Each time a new strategy, $s^*$, is added to $S^*$, it is necessarily a best response to some mixture, $m \in M(S^*)$, over the strategies already contained in $S^*$. Since strategies are never removed from $S^*$ during execution, $m$ will remain a valid mixture. Therefore, each strategy that is added is necessary to maintain the CURB property. $\square$

We will now present three algorithms that use `min_containing_CURB` to determine a game's minimal CURB sets. To facilitate our discussion of these algorithms, we first present three results regarding CURB set structure, which, to the best of our knowledge, were not previously known.

**Theorem 2.** *If each of two intersecting strategy sets is CURB, then their intersection is also CURB.*

*Proof.* Consider two CURB sets, $S_A$ and $S_B$, with nonempty intersection, $S_I$. For any mixture over strategies in $S_I$ belonging to (without loss of generality) the row player, there exists a set of pure strategies that are the column player's best responses, $S_c^*$. Because $S_A$ is CURB, it also contains all of the strategies in $S_c^*$ (i.e., $S_c^* \subseteq S_A$); likewise $S_c^* \subseteq S_B$. Therefore, $S_c^*$ is within their intersection, $S_I$. $\square$

Since the intersection of two CURB sets must be CURB and contained in both sets, we also have the following two corollaries.

**Corollary 1.** *Distinct minimal CURB sets cannot overlap (i.e., share rows or columns).*

**Corollary 2.** *Each strategy belongs to at most one minimal CURB set.*

## 4.1 Finding All Minimal CURB Sets

The broadest query one can make regarding the minimal CURB set structure of a game is asking for all of its minimal CURB sets. This is useful, for example, in the adaptive agent context to identify regions of the strategy space in which learning agents are likely to settle (Hurkens, 1995).

The `all_MC` algorithm does this by first checking each pair of strategies for size-two CURB sets (i.e., pure-strategy Nash equilibria) and adding them to the return set of minimal CURB sets. Since this operation is only $\Theta(n^2)$, it can be done as a preprocessing step without affecting the algorithm's worst-case time complexity, and the strategies it finds can be eliminated from future consideration. The `all_MC` algorithm then determines all of the minimal CURB sets in the remaining subgame by calling `min_containing_CURB` with each row strategy, in turn, as a seed.

At first, we call `min_containing_CURB` using the entire remaining subgame as an input. However, we accelerate subsequent calls to `min_containing_CURB` by maintaining a map between each strategy and the smallest CURB set in which it has been discovered so far. (The entries added to this map are also stored as candidate minimal CURB sets.) When a new strategy is used as a seed, we use the smallest known CURB set containing that strategy as the second input to `min_containing_CURB`. Whenever a smaller CURB set containing a new seed is identified, we eliminate all of the candidate minimal CURB sets that contain the newly found one. Once each strategy has been used as a seed, `all_MC` terminates and returns the remaining candidate minimal CURB sets.

**Proposition 3.** *The `all_MC` algorithm finds all of a game's minimal CURB sets, and nothing else. Its worst-case runtime is $\Theta(n^3) \times LFP(n)$. Its best-case runtime is $\Theta(n^2)$.*

*Proof.* By Corollary 1, the minimal CURB set for any strategy must either equal, or be contained by, any other CURB set in which that strategy is found. Therefore, restricting the `min_containing_CURB` search to the smallest CURB set in which a strategy has been found so far is valid, and the main loop of `all_MC` will discover all minimal CURB sets in the game. Since any CURB set that is not minimal must have contained one of the minimal CURB sets discovered, it is removed when the smaller CURB set is discovered (or not added if the smaller set was previously discovered).

In the worst case, `all_MC` must call `min_containing_CURB` $n$ times, with the full game as a parameter, giving time complexity $\Theta(n^3) \times \text{LFP}(n)$. The best-case complexity follows from a best-case game where each strategy is part of a pure-strategy Nash equilibrium. $\square$

## 4.2 Finding One Minimal CURB Set

Rather than finding all minimal CURB sets in a game, it may be desirable to find a single minimal CURB set. To complete this quickly, we first choose a random seed strategy and check if it is part of any size-two CURB sets (i.e., part of a pure-strategy Nash equilibrium), which takes $O(n)$ time. If that fails, we use the `min_containing_CURB` algorithm with the randomly-chosen strategy as the seed and the full game as the second input. Since the resulting CURB set might not be minimal, we recur within it by choosing, as a seed, a contained strategy that has not yet been used. We repeat this until all strategies in the current set have been used as seeds, at which point we terminate and return the remaining set. This constitutes the `one_MC` algorithm.

If the game has more than one CURB set, `one_MC` will be faster than `all_MC` because it will never leave the CURB set in which it starts. The exact speed of `one_MC` in practice will depend on the first seed chosen. If it happens to be in a small CURB set, `one_MC` will run fast. In the worst case, where the entire game is the only CURB set, `one_MC` executes all of the same steps as `all_MC`.

**Proposition 4.** *The `one_MC` algorithm returns one of a game's minimal CURB sets. Its worst-case time complexity is $\Theta(n^3) \times LFP(n)$. Its best-case time complexity is $\Theta(n)$.*

*Proof.* If there are no other minimal CURB sets, then the entire game is minimally CURB and will be returned. If there are any other minimal CURB sets, one of them will be discovered when a strategy inside it is used as a seed.

In the worst case, when the whole game is minimally CURB, `one_MC` must call the `min_containing_CURB` algorithm $n$ times, with the full game as an input, giving time complexity $\Theta(n^3) \times \text{LFP}(n)$. The best-case complexity follows from a best-case game where a strategy that is in a CURB set of size two is chosen as a seed. $\square$

### 4.3 Finding the Smallest Minimal CURB Set

As a different type of query, one may be interested in finding one of a game's *smallest* minimal CURB sets. This is important, for example, if the CURB set is used for future computations and the complexity of those future computations increases with the size of the CURB set (e.g., for Nash equilibrium finding as we will discuss later in the paper). We find one of a game's smallest minimal CURB sets using a pseudo-parallelization of `all_MC`.

First, we use the same preprocessor as in `all_MC` that checks each pair of strategies for a size-two CURB set and returns one, if found. If that fails, we construct a candidate set for each row strategy containing only that strategy. We insert the sets into a priority queue, where sets containing the fewest strategies are given highest priority. We repeatedly pop the smallest candidate set from the queue, and add all the necessary best responses to keep it CURB by calling `all_conditionally_rational` for each player. If new strategies are added for either player, the resulting set is inserted back into the queue, and it is prioritized based on its new size. The algorithm terminates when a candidate set is removed from the queue that fails to admit any new best responses. That set is returned and it is one of the game's smallest minimal CURB sets (we denote the size of this set as $n^*$). We call this algorithm `small_MC`.

**Proposition 5.** *The* `small_MC` *algorithm returns one of the game's smallest minimal CURB sets. Its worst-case runtime is* $\Theta(n^* n^2) \times LFP(n)$. *Its best-case runtime is* $\Theta(n^2)$.

*Proof.* At the time `small_MC` terminates, `all_conditionally_rational` has been called on each row and column strategy in the set with no new best responses having been added. Therefore, the returned set is CURB. Since all other candidate sets on the queue must be as large, or larger than the returned set (and future exploration can only add strategies to these sets), this set is at least as small as the smallest CURB set in the game, and each of the game's smallest CURB sets is also minimal.

Whenever a candidate set is fathomed, at least one new strategy must be added or `small_MC` will terminate. Since there are $n$ candidate sets, and $n^*$ strategies in the returned set, in the worst case $n \times n^*$ sets have been fathomed at termination. Since each examination of a candidate set involves a call to `all_conditionally_rational`, the complexity of `small_MC` is as claimed. Priority queue operations are logarithmic in the size of the game, and in the worst case there are $n \times n^*$ such operations. Thus, the overall worst-case complexity is $\Theta(n^* n^2 + n^* n \log n) \times \text{LFP}(n)$, which is $\Theta(n^* n^2) \times \text{LFP}(n)$. The proof of the best-case complexity is the same as in Proposition 3. $\square$

### 4.4 Experimental Results

We implemented the algorithms above and conducted experiments on their performance using most of the instance generators of the main benchmark collection for solving normal-form games, *GAMUT* (Nudelman, Wortman, Shoham, & Leyton-Brown, 2004). The *GAMUT* collection includes a variety of commonly studied game types from the academic game theory literature. It is also specifically designed to test different aspects of scalability for game-solving algorithms, for example, most of the generators allow one to create multiple game instances of any given size.[1] In this section we show that the complexity of our algorithms depends primarily on the size of the game and the size of its smallest CURB set. We then proceed to explore the distribution of CURB set sizes in the different game types.

We first report the runtime of our algorithms on two representative *GAMUT* game distributions: *random games*, and *covariant games*. Figure 1 (left) shows how each of our minimal CURB set finding algorithms scales with game size on a data set of over 1000 random, square normal-form games with between 20 and 100 strategies. The results show that `small_MC` is faster than `all_MC` on random games, which is consistent with their time complexities, considering that many random games have small CURB sets. While the worst-case time complexity of `one_MC` and `all_MC` is the same, experimentally `one_MC` is faster because it only needs to find one minimal CURB set. We can also see that, for large random games, `small_MC` performs slightly better than `one_MC`, since these games tend to contain both small and large CURB sets and `one_MC` is more likely to start in the larger ones. On the other hand, in games with only large CURB sets, `one_MC` tends to be faster, as we will show later.

We observed that the performance on random games, illustrated in Figure 1 (left), was typical of that of many other *GAMUT* instance distributions. However, to show potentially differing performance, we also report on experiments with the covariant game class, in which utilities for both players are drawn from the same distribution with a specified covariance. (In our experiments we set the covariance parameter to be $-0.5$.) This class (and setting) have been shown to be particularly challenging for Nash equilibrium finding algorithms, such as the Lemke-Howson and Porter-Nudelman-Shoham algorithms (Lemke & Howson, 1964; Porter et al., 2004). Figure 1 (right) shows that the `all_MC` algorithm scales similarly on random and covariant games, while the other two algorithms lose their speed advantages when applied to this class.

The distribution of CURB sets in random games is shown as solid dots in Figure 2. Most random games have small smallest CURB sets (in fact, often sets of size two), and those that do not, tend to have very large smallest CURB sets. On the other hand, the distribution of smallest CURB set sizes in covariant games (shown in Figure 2, hollow squares) has almost no small smallest CURB sets and many large smallest CURB sets. This is consistent with the observed hardness of these games for support enumeration-based Nash equilibrium finding algorithms, which typically try to find equilibria with small supports first (Porter et al., 2004). This disparity also explains the lowered performance on covariant games of the two minimal CURB finding algorithms that have time complexities dependent on the size of the smallest minimal CURB set, `small_MC` and `one_MC`.

Figure 3 shows the distribution of smallest CURB set size for 1000 instances from each of the twenty-four other distributions emitted by *GAMUT* generators. Using a variety of game generators, as we have done here, has become a primary way of testing game-solving algorithms (Porter et al., 2004; Sandholm et al., 2005), and we used the same parameter settings in the distributions as those prior papers. For covariant games, the suffixes "Pos", "Rand", and "Zero" refer to positive, random, and zero covariance parameters, respectively. For distributions that take a graph as input, "CG", "RG", and "SG" refer to complete, random, and star graphs.

---

1. We did not benchmark on the *GAMUT* games that only have a fixed size, such as Chicken, Prisoner's Dilemma, and Battle of the Sexes, because they are trivial to solve from a computational perspective.

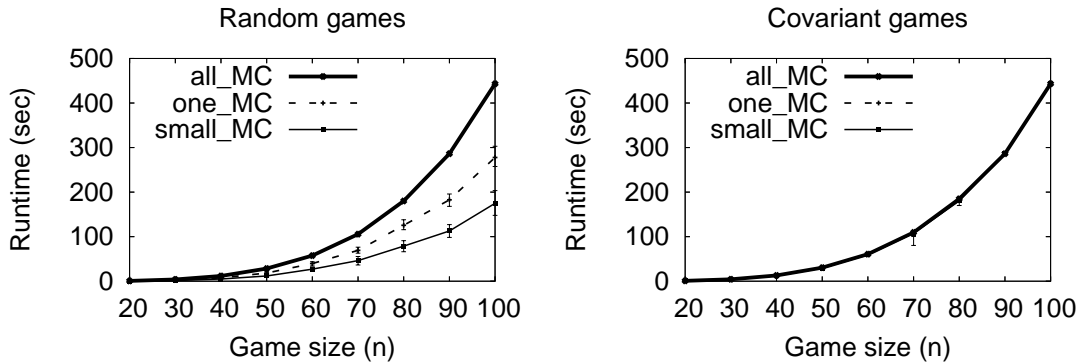

Figure 1: Scalability of our algorithms in game size for random (left) and covariant (right) games.

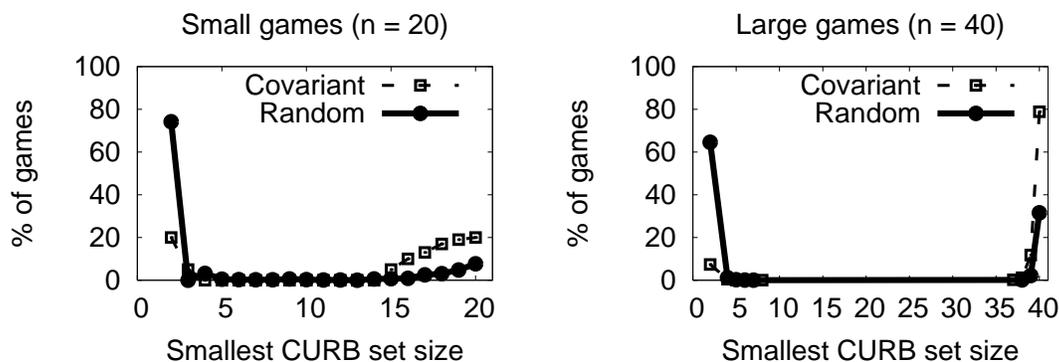

Figure 2: Distribution of smallest CURB set size in random and covariant ($r = -0.5$) games, where $n = 20$ and $n = 40$ (3,000 games were generated for each distribution and value of $n$).

All of these distributions, like random and covariant games, exhibited very few medium-sized smallest CURB sets. Most of the instances had a smallest CURB set that was extreme: either a pure strategy equilibrium or the entire game. With some of the generators, all of the instances lie at the same extreme. Interestingly, some generators (e.g., Polymatrix) produced a significant number of games with CURB sets of one or more specific, non-extremal sizes. It is also notable that using different graph parameters for Local Effect and Polymatrix games had no effect on their smallest CURB set size distributions, suggesting

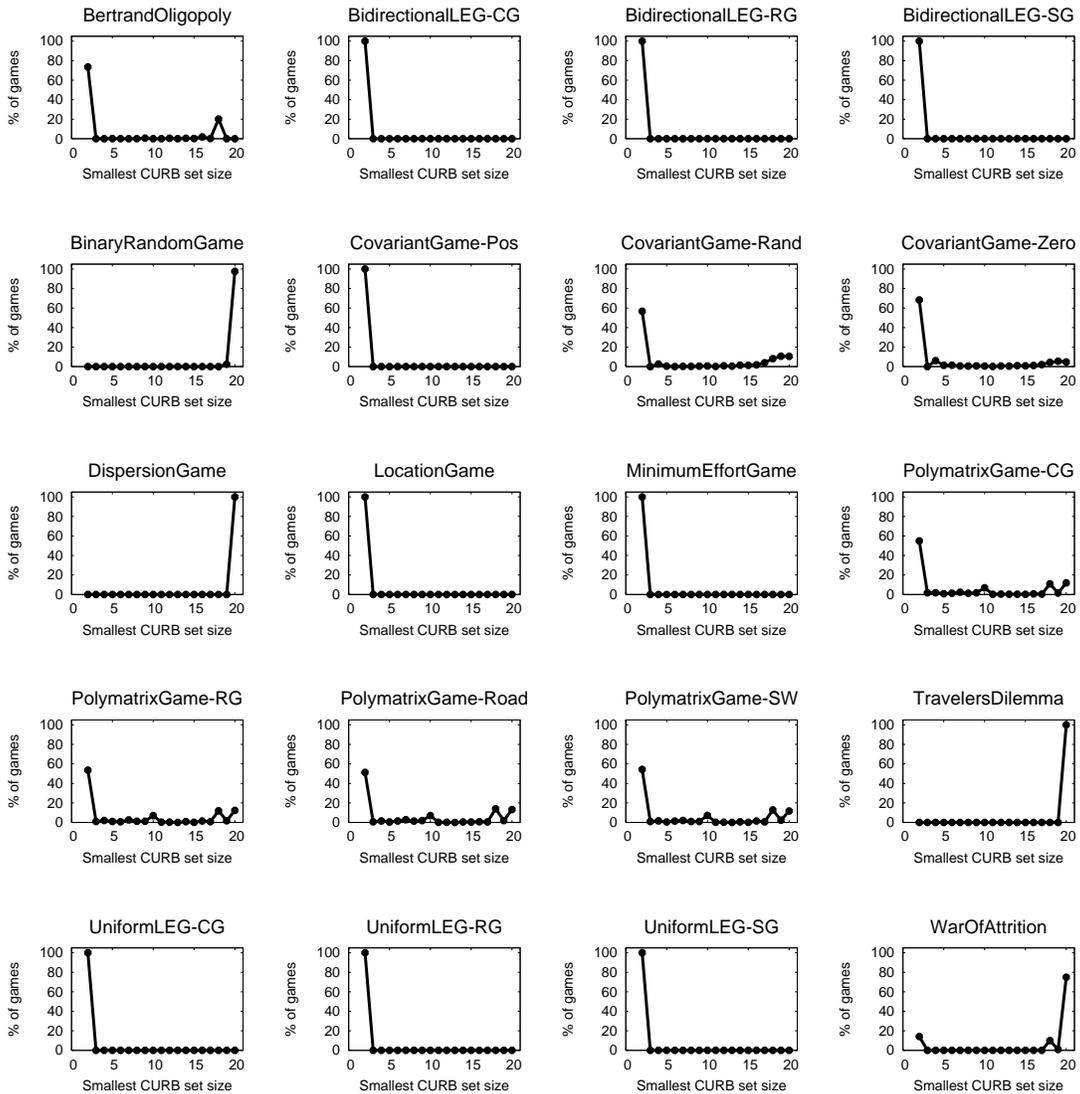

Figure 3: Distribution of smallest CURB set size in sets of 1000 games with $n = 20$ from various *GAMUT* distributions.

that the type of graph used may not change the fundamental structure of these types of games.

To better understand how the minimal CURB set finding algorithms scale with the size of the smallest CURB set, we bucketed the $n = 20$ random and covariant games according to the size of their smallest CURB sets. (For $n = 40$, the buckets for medium-sized smallest CURB sets were nearly empty, making it impossible for us to estimate mean runtimes with meaningful accuracy.) Figure 4 plots the average runtime and 95% confidence intervals for each bucket. On games with very small CURB sets, `small_MC` is fastest, but it is outperformed by both `one_MC` and `all_MC` as the size of the smallest CURB set grows.

The somewhat surprising runtime performance of the latter two algorithms is due to their leveraging of information across calls to `min_containing_CURB` with different seeds. Because `small_MC` performs all the searches in parallel, this information is unavailable.

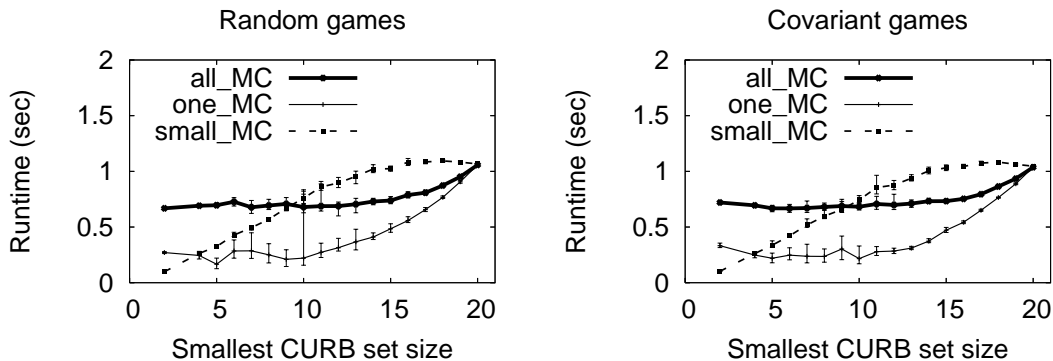

Figure 4: Average runtime on games were $n = 20$, with varying smallest CURB set sizes.

## 5. CURB Sets and Nash Equilibria

Minimal CURB sets and Nash equilibria both model strategy subspaces which are mutually reinforced given the rationality of agents and their common knowledge. In their original work on minimal CURB sets, Basu and Weibull showed that every minimal CURB set contains the supports of at least one Nash equilibrium (Basu & Weibull, 1991). We observe that this result suggests a secondary use for finding minimal CURB sets: our algorithms can be used to preprocess a game so that a Nash equilibrium finding algorithm can restrict its attention to one of the game's minimal CURB sets, rather than running on the entire game. As we will show, this can theoretically yield an arbitrarily large reduction of the search space.

The most common prior preprocessing technique for Nash equilibrium finding, iterated removal of dominated strategies, attempts to eliminate strategies that cannot be played with any probability in any Nash equilibrium (Knuth et al., 1988; Gilboa, Kalai, & Zemel, 1993). The same is true of another recent preprocessing technique, the generalized eliminability method (Conitzer & Sandholm, 2005b). One comparative advantage of minimal CURB set-based elimination is that it can eliminate strategies that are played in some equilibria, while guaranteeing that the resulting set still contains the supports of at least one.

First, we will show that a CURB set-based preprocessor can reduce search space size by an arbitrary amount even on games where prior preprocessing techniques cannot eliminate anything.

**Theorem 3.** *For any $r,c,r'$, and $c'$ such that $r \geq 2$, $c \geq 2$, $1 < r' \leq r$, and $1 < c' \leq c$, there exists normal form games of size $r \times c$, with the following properties:*

*a) it contains a minimal CURB set with shape $r' \times c'$,*

b) *iterated elimination of dominated strategies (even domination by mixed strategies) cannot eliminate any strategies,*

c) *the recursive preprocessing technique (that **can** eliminate strategies that belong to some equilibrium as long as some other equilibrium remains) (Conitzer & Sandholm, 2006) can eliminate at most one strategy per player, and*

d) *the general eliminability method (Conitzer & Sandholm, 2005b) cannot eliminate any strategies.*

*Proof.* We first present a family of games, $\Gamma$. Let $\Gamma_{r'c'}$ denote a game from this family of size $r' \times c'$. The following generator produces such a game where $r', c' \geq 2$. Assign the utilities,

- $u(s_{r_1}, s_{c_1}) = u(s_{r_2}, s_{c_2}) = (0, 1)$ and $u(s_{r_1}, s_{c_2}) = u(s_{r_2}, s_{c_1}) = (1, 0)$.

Then, for $i \in [2, \lfloor \frac{r'}{2} \rfloor]$, set

- $u(s_{r_{2i-1}}, s_{c_1}) = (\frac{r'-2i+2}{r'}, 1)$ and $u(s_{r_{2i-1}}, s_{c_2}) = (\frac{2i-2}{r'}, 0)$,

- $u(s_{r_{2i}}, s_{c_1}) = (\frac{r'-(2i-1)}{r'}, 0)$ and $u(s_{r_{2i}}, s_{c_2}) = (\frac{2i-1}{r'}, 1)$.

If $r'$ is odd there will be one remaining row. In this case, set the following utilities,

- if $r'$ is odd, $u(s_{r'}, s_{c_1}) = (\frac{1}{r'}, \frac{1}{2})$ and $u(s_{r'}, s_{c_2}) = (\frac{r'-1}{r'}, \frac{1}{2})$.

Next, for $j \in [2, \lfloor \frac{c'}{2} \rfloor]$, set

- $u(s_{r_1}, s_{c_{2j-1}}) = (0, \frac{c'-2j+2}{c'})$ and $u(s_{r_2}, s_{c_{2j-1}}) = (1, \frac{2j-2}{c'})$,

- $u(s_{r_1}, s_{c_{2j}}) = (1, \frac{c'-(2j-1)}{c'})$ and $u(s_{r_2}, s_{c_{2j}}) = (0, \frac{2j-1}{c'})$.

If $c'$ is odd there will be one remaining column. In this case, set the following utilities,

- if $c'$ is odd, $u(s_{r_1}, s_{c'}) = (\frac{1}{2}, \frac{1}{c'})$ and $u(s_{r_2}, s_{c'}) = (\frac{1}{2}, \frac{c'-1}{c'})$.

For example, the game $\Gamma_{3,4}$ is as follows.

| $\mathbf{\Gamma_{3,4}}$ | $s_{c_1}$ | $s_{c_2}$ | $s_{c_3}$ | $s_{c_4}$ |
|---|---|---|---|---|
| $s_{r_1}$ | 0,1 | 1,0 | $0,\frac{1}{2}$ | $1,\frac{1}{4}$ |
| $s_{r_2}$ | 1,0 | 0,1 | $1,\frac{1}{2}$ | $0,\frac{3}{4}$ |
| $s_{r_3}$ | $\frac{1}{3},\frac{1}{2}$ | $\frac{2}{3},\frac{1}{2}$ | -3,-3 | -3,-4 |

Any game generated in this way has a Nash equilibrium where the row player mixes evenly between its first two strategies, and the column player mixes evenly among all of its strategies. This game also has an equilibrium where the column player mixes evenly between its first two strategies, and the row player mixes evenly among all of its strategies. Thus, every strategy in $\Gamma_{r'c'}$ is part of some equilibrium. Additionally, each column strategy is a best response to a mixture over the first two row strategies (and, to any column strategy, one of those two is a best response), and vice-versa. Thus, $\Gamma_{r'c'}$ has a single minimal CURB set and it includes the entire game.

We now construct an $r \times c$ game, with a minimally CURB $r' \times c'$ subset, by putting the game $\Gamma_{r'c'}$ in the top left and the game $\Gamma_{(r-r')(c-c')}$ in the bottom right. All other utilities are set to arbitrary negative values, such that no two are exactly the same. The resulting game is shown in Figure 5.

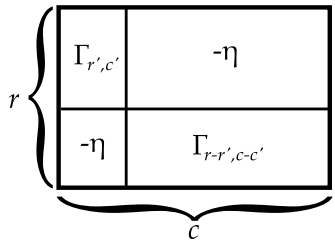

Figure 5: $r \times c$ game with arbitrary reduction to an $r' \times c'$ CURB set, irreducible by prior techniques.

It is irreducible by (iterated) dominance and by general eliminability because every strategy participates in some Nash equilibrium. The game is irreducible (other than a single strategy per player) by the recursive preprocessor because the row player's utilities are distinct within each column, and the column player's utilities are distinct within each row (except for the last row or column in each $\Gamma$ game, if it has an odd number of rows or columns). □

Three factors curb the promise of minimal CURB set algorithms as powerful preprocessors for Nash equilibrium finding. First, the fastest Nash equilibrium finding algorithms, while requiring exponential time in the worst case, tend to run faster than the CURB set finding algorithms on many types of games (at least the best known implementations of these algorithms). Second, the smallest CURB set can be arbitrarily large (up to the size of the entire game, in which case the preprocessor does not eliminate any strategies from consideration). Third, as we will now show, even after the smallest minimal CURB set has been identified, the remaining search space (CURB set size) can be arbitrarily larger than the size of the supports of a contained Nash equilibrium.

To prove this, we will use the following family of games that contain large minimal CURB sets and small-support equilibria. For any integer $k > 0$, we define the game $\Omega_k$ as follows. As in the previous proof, assign the utilities $u(s_{r_1}, s_{c_1}) = u(s_{r_2}, s_{c_2}) = (0, 1)$ and $u(s_{r_1}, s_{c_2}) = u(s_{r_2}, s_{c_1}) = (1, 0)$, and let $Z$ be an arbitrarily large value (essentially $\infty$). Then, for $i \in [3, 2 + k]$,

- $u(s_{r_i}, s_{c_1}) = (-Z, \epsilon)$, $u(s_{r_1}, s_{c_i}) = (\epsilon, -Z)$,

- $u(s_{r_i}, s_{c_i}) = (0, 0)$,

- $u(s_{r_i}, s_{c_{i-1}}) = (1 + \epsilon, 0)$, $u(s_{r_{i-1}}, s_{c_i}) = (0, 1 + \epsilon)$, and

- for all $j > i + 1$ and $j \leq 2 + n$,

$$u(s_{r_i}, s_{c_j}) = (0, -Z), \text{ and } u(s_{r_j}, s_{c_i}) = (-Z, 0)$$

For example, the game $\Omega_2$ is as follows.

| $\Omega_2$ | $s_{c_1}$ | $s_{c_2}$ | $s_{c_3}$ | $s_{c_4}$ |
|---|---|---|---|---|
| $s_{r_1}$ | 0,1 | 1,0 | $\epsilon,-Z$ | $\epsilon,-Z$ |
| $s_{r_2}$ | 1,0 | 0,1 | $0,1+\epsilon$ | $0,-Z$ |
| $s_{r_3}$ | $-Z,\epsilon$ | $1+\epsilon,0$ | 0,0 | $0,1+\epsilon$ |
| $s_{r_4}$ | $-Z,\epsilon$ | $-Z,0$ | $1+\epsilon,0$ | 0,0 |

With respect to the strategic structure of games from this class, we have the following results.

**Lemma 1.** *For $i > 2$, the row (column) player's strategy $s_{r_i}$ $(s_{c_i})$ is a best response to the column (row) player's strategy $s_{c_{i-1}}$ $(s_{r_{i-1}})$. The column (row) player's strategy $s_{c_1}$ $(s_{r_1})$ is a best response to the row (column) player's strategy $s_{r_{n+2}}$ $(s_{c_{n+2}})$.*

**Proposition 6.** *$\Omega_k$ has a single minimal CURB set and it includes the entire game.*

*Proof.* Strategies $s_{r_1}, s_{r_2}, s_{c_1}$, and $s_{c_2}$ must be included in some minimal CURB set, as they are the best responses to each other in the subgame containing them, and this subgame admits no pure-strategy Nash equilibrium. Based on Lemma 1, we can see that when $i = 3$, the row (column) player's strategy $s_{r_3}$ $(s_{c_3})$ is a best response to the column (row) player's second strategy. This forces the third strategy of each player into the minimal CURB set containing the first two strategies of each player, and inductively each additional strategy is added in the same way. $\square$

**Proposition 7.** *In $\Omega_k$, the only Nash equilibrium is the the mixed-strategy profile where $s_{r_1}$, $s_{r_2}$, $s_{c_1}$, and $s_{c_2}$ are each played with probability $\frac{1}{2}$.*

*Proof.* Assume, for contradiction, that this is not the case, that is, there exists a mixture, $m_r^*$, over the rows $M_r^*$, comprising the row player's profile in a Nash equilibrium, and $s_{r_1} \notin M_r^*$. Along with our assumption, the definition of Nash equilibrium implies that there must exist a mixture, $m_c^*$, over columns $M_c^*$ such that $\beta_r(m_c^*) = M_r^*$ and $\beta_c(m_r^*) = M_c^*$. Since $s_{r_1}$ is not in $M_r^*$ by assumption, there exists $i > 1$ such that $s_{r_i}$ is the lowest numbered support in $M_r^*$, and the definition of $\Omega$ specifies the outcome, $u(s_{r_i}, s_{c_j}) = (0, -Z)$, when $j > i + 1$.

The column player's Nash equilibrium supports cannot contain any such $s_{c_j}$, because placing any positive probability on this strategy will lead to a highly negative expected payoff and playing the pure strategy $s_{c_1}$ provides guaranteed payout of at least 0. If we exclude these strategies, $s_{c_{i+1}}$ (the highest-numbered remaining column strategy) is the only remaining strategy, other than $s_{c_1}$, which provides non-zero utility against mixtures on rows $\geq i$. In other words, it dominates all column strategies on the row player's supports $M_r^*$, aside from one: $s_{c_1}$.

Since dominated strategies cannot be played in equilibrium, $M_c^*$ is constrained to a subset of $\{s_{c_1}, s_{c_{i+1}}\}$. If $M_c^*$ contains $s_{c_1}$, $M_r^*$ must not include any $s_{r_j}$ with $j > 2$, due to the highly negative expected payoff of any mixture including such strategies (as discussed above). In this case, the only remaining possible equilibrium row mixture is the pure strategy $s_{r_2}$, the best response to which is $s_{c_3}$. Since, by Corollary 2, $\Omega^n$ has no pure Nash equilibrium, this cannot constitute an equilibrium, contradicting our assumption. Alternatively, if $M_c^*$ does not include $s_{c_1}$, then $m_c^*$ must be the pure strategy $s_{c_{i+1}}$, and Lemma 1 provides a pure-strategy best response to any pure strategy with $i > 2$. This would, again, form a pure strategy Nash equilibrium, which we have shown cannot exist.

The above reasoning can also be inverted to show that a contradiction is caused by the assumption that $M_c^*$ does not contain $s_{c_1}$.

We have shown that the row player's equilibrium mixture must contain $r_1$ and $r_2$, and that the column player's equilibrium mixture must contain $c_1$ and $c_2$ and no other strategies can be in either player's supports, since it would lead one of them to have to highly negative utility. □

The $\Omega$ game demonstrates that it is possible to construct very large CURB sets that are loose around the supports of an enclosed Nash equilibrium, giving us the following general result.

**Theorem 4.** *A Nash equilibrium with supports consisting of two strategies for each player can be the only Nash equilibrium in an arbitrarily large minimal CURB set.*

These results imply that minimal CURB set algorithms will not always be effective as preprocessors for Nash equilibrium finding. However, on game instances that have a small CURB set *or* a relatively tight minimal CURB set,[2] these algorithms have the potential to yield a significant speed improvement.

Furthermore, the existence of the polynomial-time algorithm for detecting a game's smallest CURB set (`small_MC`) allows us to offer the following theoretical result of potential general interest.

**Theorem 5.** *The complexity of finding a Nash equilibrium for a two-player normal-form game can be super-polynomial only in the size of the game's smallest CURB set (not in the size of the full game).*

The relationship between the complexity of finding a minimal CURB set and that of finding a Nash equilibrium is surprising in several ways. For one, it is not obvious that finding a minimal CURB set should be easier than finding a Nash equilibrium, since, like Nash equilibria, CURB sets have an exponential space of possible supports which are chosen through maximization processes for both players. Yet from a theoretical, worst-case perspective, Nash equilibrium finding is PPAD-complete (which is widely believed to be a strictly harder complexity class than P) and, as we showed earlier in this paper, minimal CURB set finding is polynomial time.

It is worth noting that games with very small support equilibria, which include all games with very small CURB sets, are already known to be easily solvable for Nash equilibria using techniques such as support enumeration. In particular, on games whose smallest CURB set size is logarithmic in the full game size, both support enumeration and CURB set preprocessing permit a guarantee of polynomial runtime in finding a Nash equilibrium. CURB set preprocessing has the additional advantage that it can also be used to simplify games with larger equilibrium supports, where support enumeration is exponential. For example, consider the $G_k$ game class described by Sandholm, Gilpin and Conitzer (2005), which generates games with a single equilibrium, and that equilibrium contains half the strategies in its support. We also determined that these games have a single CURB set, and that CURB set includes exactly the supports of the equilibrium. Games from this class can be padded, using the embedding technique in our Theorem 3, to become arbitrarily large games without introducing any additional equilibria or CURB sets. On those games, CURB set detection offers a polynomial-time method for reducing the game to the point where algorithms not based on support enumeration can be applied.

---

2. If the game has a relatively tight CURB set, a Nash equilibrium can be found quickly by enumerating strategies of the CURB to be left out from the supports.

What about the complexity of the two problems (Nash equilibrium finding and CURB set finding) in practice? As Figure 6 shows, the average runtimes of our smallest CURB set finding algorithm and the Lemke-Howson Nash equilibrium finding algorithm (using the implementation in *Gambit*, McKelvey, McLennan, & Turocy, 2004) seem to scale similarly with input game size (when one does not explicitly generate those pathological cases which produce exponential behavior in the latter, Savani & von Stengel, 2004). In fact, the CURB set algorithms are *slower* (by two orders of magnitude) on average than Lemke-Howson. This experimental performance agrees with intuition, but is the reverse of the theoretical state of knowledge regarding worst-case complexity.

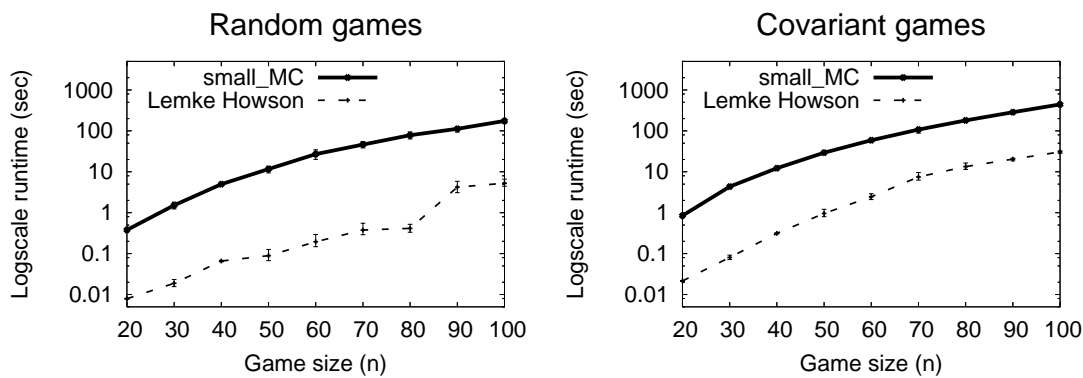

Figure 6: Average runtime and 95% confidence intervals of `small_MC` and Lemke Howson as a function of game size.

It is worth pointing out that an algorithm that builds in part on our work for one of the CURB set problems (finding all minimal CURB sets) has been presented in a working paper (Klimm, Sandholm, & Weibull, 2010), and it appears to scale more favorably than ours. However, the algorithm has not been directly compared with Lemke-Howson or any other Nash equilibrium finding algorithms, so its relative value related to preprocessing remains to be seen.

The root cause of the complexity of Nash equilibrium search has proved elusive, as two candidates that were considered potential culprits have been shown not to affect worst-case complexity: games with two players and binary utilities are just as difficult as the general case, even if both restrictions apply simultaneously (Chen & Deng, 2006; Abbott et al., 2005). The fact that bounding the smallest CURB set size *does* serve to bound the difficulty of Nash equilibrium search suggests that we can further isolate the cause of equilibrium search complexity as being endemic to minimal CURB sets, rather than games in general. In this regard, we observe that the special two-player game used by Chen and Deng to show PPAD-completeness (Chen & Deng, 2006) is itself a single minimal CURB set, and remains such under Abbott *et al.*'s (Abbott et al., 2005) transformation to binary utilities.

## 6. Conclusions

We presented a thorough computational treatment of CURB sets, an important set-valued, game-theoretic solution concept, including several algorithms for finding CURB sets in two-player, normal-form games. Our algorithms find all minimal CURB sets (`all_MC`), one minimal CURB set (`one_MC`), and a smallest minimal CURB set (`small_MC`), all in polynomial time. The algorithms are based on basic properties of CURB sets that we prove, such as the fact that minimal CURB sets cannot overlap. The algorithms use dovetailing with a priority queue, and exploiting information across overlapping, non-minimal CURB sets, to further improve speed.

Experiments on random games showed that, unsurprisingly, `small_MC` tends to be the fastest, `one_MC` is second, and `all_MC` is the slowest. However, on covariant games the speed advantage of the former two disappears. The runtime of those algorithms is primarily determined by the size of the smallest CURB set, and on covariant games, which tend to have larger CURB sets, those algorithms (especially `one_MC`) suffer.

Our algorithms also enable the study of CURB set size distributions of different game classes. We showed that the instance distributions from *GAMUT* are mainly extremal, in the sense that a given game generator will yield mostly games with pure-strategy equilibria and/or games where the game itself is the sole minimal CURB set. However, curiously, some of the generators yield a significant number of games with smallest CURB sets of specific non-extremal sizes.

We also examined the potential for using our algorithms as preprocessors for Nash equilibrium finding algorithms. We proved that our technique can eliminate an arbitrarily large portion of the game from consideration, while guaranteeing that the remaining subgame contains at least one Nash equilibrium from the full game. This is the case even for games where prior preprocessing techniques, including the iterated removal of dominated strategies, are powerless.

On the downside, we showed that the smallest CURB set can be arbitrarily large and/or arbitrarily loose. Furthermore, on many distributions, we showed that current Nash equilibrium finding algorithms run faster, on average, than the CURB set algorithms. This is surprising in that the theoretical worst-case complexity of the two problems is the reverse.

We demonstrated that the worst-case complexity of finding a Nash equilibrium is polynomial in all known aspects of the game *except* the size of its smallest CURB set. Taken together with our CURB set finding algorithms that are polynomial time even in the worst case, and the fact that Nash equilibrium finding is super-polynomial in the worst case (unless PPAD=P), we observe that the essence of the worst-case complexity of finding a Nash equilibrium is the complexity of finding a Nash equilibrium within a minimal CURB set.

While the CURB set definition is for any number of players, we presented all of our algorithms for the two-player setting. For a larger number of players, the only obstacle to finding minimal CURB sets is finding conditional best responses quickly as a subroutine. We showed that this problem can be solved fast for two players—in these settings it involves solving a simple linear feasibility program. However, the mathematical program we use is of degree $p-1$, where $p$ is the number of players, and with three players the constraints are already quadratic. On the plus side, our algorithms only make a polynomial number of calls to this subroutine. Therefore, if future research is able to identify polynomial-time algorithms for finding all of a player's conditional best responses in $n$-player games, then our CURB set algorithms will also be polynomial time in those settings.

## Acknowledgments

This material is based upon work supported by National Science Foundation ITR grant 0205435, IGERT grant 9972762, and IIS grants 0121678, 0427858, and 0905390, as well as Office of Naval Research grant N00014-02-1-0973, and a Sloan Fellowship. We would also like to thank our anonymous reviewers, Vincent Conitzer, and Andrew Gilpin for their helpful input and advice.